\documentclass[aps,prd,preprint,floats,epsf,superscriptaddress,nofootinbib]{revtex4}

\usepackage[utf8]{inputenc}
\DeclareUnicodeCharacter{2212}{\-}
\usepackage{float}
\usepackage{subcaption}
\usepackage{graphicx} 
\usepackage{amsmath,amssymb,amsfonts,mathtools} 
\usepackage{mathrsfs}
\usepackage{slashed} 
\usepackage{braket} 
\usepackage{indentfirst}
\usepackage{xcolor}
\usepackage{dsfont}
\usepackage[normalem]{ulem}
\usepackage{multirow}
\usepackage{cancel}
\usepackage{dcolumn}
\usepackage{bm}
\usepackage{bbm}
\usepackage{tabularx}

\usepackage[hyperfootnotes=true]{hyperref}
\allowdisplaybreaks

\graphicspath{{Plots/}}

\begin{document}
\title{ VBF vs.\ GGF Higgs with Full-Event Deep Learning: \\ Towards a Decay-Agnostic Tagger}

\author{Cheng-Wei Chiang} 
\email[email: ]{chengwei@phys.ntu.edu.tw}
\affiliation{Department of Physics, National Taiwan University, Taipei, Taiwan 10617, ROC}
\affiliation{Physics Division, National Center for Theoretical Sciences, Taipei, Taiwan 10617, ROC}

\author{David Shih} 
\email[email: ]{shih@physics.rutgers.edu}
\affiliation{NHETC, Department of Physics and Astronomy, Rutgers University, NJ 08854, USA}

\author{Shang-Fu Wei}
\email[email: ]{b06202061@ntu.edu.tw}
\affiliation{Department of Physics, National Taiwan University, Taipei, Taiwan 10617, ROC}

\vspace*{1cm}
\begin{abstract}
We study the benefits of jet- and event-level deep learning methods in distinguishing vector boson fusion (VBF) from gluon-gluon fusion (GGF) Higgs production at the LHC. We show that a variety of classifiers (CNNs, attention-based networks) trained on the complete low-level inputs of the full event achieve significant performance gains over 
shallow machine learning methods (BDTs) trained on jet kinematics and jet shapes, and we elucidate the reasons for these performance gains. Finally, we take initial steps towards the possibility of a VBF vs.\ GGF tagger that is agnostic to the Higgs decay mode, by demonstrating that the performance of our event-level CNN does not change when the Higgs decay products are removed. These results highlight the potentially powerful benefits of event-level deep learning at the LHC.
\end{abstract}

\maketitle

\section{Introduction}

The discovery~\cite{ATLAS:2012yve,CMS:2012qbp} of the Higgs boson in 2012 was a monumental occasion, providing a capstone to decades of experimental and theoretical works in particle physics, and confirming the final missing piece of the Standard Model (SM).

Since the original discovery, much effort~\cite{Cepeda:2019klc,grazzini_2019} has been devoted to measuring ever more precisely the couplings of the Higgs boson to other SM particles.  Since the Higgs has numerous production modes and decay modes, measurements in many different final states are necessary to disentangle all the various effects and pin down the Higgs couplings to all the SM fields~\cite{ATLAS:2020qdt,ATLAS:2020rej,ATLAS:2020pvn,ATLAS:2021upe,CMS:2021kom,CMS:2022xam,CMS:2021ugl}. A key component of this program is distinguishing the vector boson fusion (VBF) production mode from other production modes, most predominantly gluon-gluon fusion (GGF). VBF is essential for measuring the Higgs couplings to the SM $W/Z$ gauge bosons, thereby testing the most essential property of the Higgs, namely its role in electroweak symmetry breaking (EWSB).

Previous works~\cite{Chan:2017zpd,Chung:2020ysf} have studied the question of VBF vs GGF classification with machine learning methods (see also~\cite{Rentala:2013uaa} for earlier work on high-level features for VBF vs GGF). The main thing that distinguishes VBF from GGF events is that VBF events come with two forward quark-initiated jets from the hard process, while GGF jets are going to be from ISR and will tend to be gluon-initiated. 
In Ref.~\cite{Chan:2017zpd}, boosted decision trees (BDTs) trained on high-level physics variables such as invariant mass and rapidity difference of the leading jets, sum of transverse momenta of the Higgs decay products, and various jet shape variables were brought to bear on the question of VBF vs.\ GGF classification, in the context of $H\to\gamma\gamma$ and $H\to W W^*$ final states specifically. Meanwhile, Ref.~\cite{Chung:2020ysf} studied the multiclass classification of multiple Higgs production modes (including VBF and GGF) in the boosted $H\to bb$ regime, considering BDTs trained on high-level features, as well as a specialized two-stream convolutional neural network (CNN), which was previously developed for boosted $H\to bb$ tagging~\cite{Lin:2018cin}, and was trained on event images made out of low-level inputs (the pixelated $p_T$'s of all the particles in the event).
 
Experimental studies~\cite{ATLAS:2021upe,ATLAS:2020pvn,ATLAS:2020qdt,ATLAS:2020rej,ATLAS:2021pkb,ATLAS:2020fzp,CMS:2021kom,CMS:2022ahq,CMS:2022xam,CMS:2020xrn} have also used BDTs or  dense neural networks (DNNs) on a variety of Higgs decay modes to discriminate VBF from GGF events, while other techniques such as recurrent neural networks (RNNs) were also found useful in practice~\cite{ATLAS:2020rej}. The BDTs, DNNs, RNNs used by the experimental groups take the high-level features as input.

In this work we will revisit the question of VBF vs GGF event-level classification, exploring the benefits that machine learning methods (both shallow and deep) can bring to this problem. Our starting point will be a BDT trained on high-level features (HLFs) defined from the leading two jets and the Higgs decay products; this baseline method is designed to characterize the previous state of the art from \cite{Chan:2017zpd} and from the actual ATLAS and CMS analyses. To go beyond, we consider the following methods:
\begin{itemize}

\item Training a jet-level CNN to distinguish the leading two jets from VBF from their GGF counterparts, and adding the jet-CNN scores to the inputs of the HLF BDT.

\item Training an event-level CNN to distinguish full VBF events from full GGF events; we make full-event images out of the energy deposits of all the reconstructed particles in the event.

\item Training an event-level network based on the self-attention mechanism~\cite{DBLP:journals/corr/LinFSYXZB17,https://doi.org/10.48550/arxiv.1706.03762} as an interesting alternative to the event-level CNN. In such a self-attention model, we convert the input event into a sequence which directly records the detector-level information. 

\end{itemize}

We will see that while augmenting the HLFs with the jet-CNN scores offers some gain in classification performance, a much bigger boost comes from the event-level classifiers trained on low-level inputs. We investigate the reasons for the performance gains of the event-level CNN and find it is due in part to additional hadronic activity beyond the leading two jets. Interestingly, this includes both additional jet activity, as well as unclustered hadronic activity in the event (i.e., hadronic activity that leads to softer jets below the jet $p_T$ threshold). The pattern of soft radiation is different in VBF vs.\ GGF events, again presumably due to differing quark vs.\ gluon content in the initial states. 

In this paper we will also highlight an added benefit of event-level classifiers trained on low-level inputs: they can be {\it Higgs decay mode agnostic.} Since the Higgs is a color singlet, the Higgs decay should be fairly well factorized from the VBF or GGF initial state jets, especially when it decays to electroweak states. Besides, the $p_T$-balance of the full event ensures that the kinematics of the Higgs can be well-reconstructed from all the other final state objects. Using the diphoton mode as an explicit example, we will show that as long as our models take the whole event into account,  adding information from the Higgs decay does not improve the performance of the classifier. This raises the possibility that a single VBF vs.\ GGF classifier could be trained and deployed in a variety of Higgs analyses with different final states, with no loss in performance. 

Much work in the literature has focused on boosted jet classification~\cite{Pumplin:1991kc,Cogan:2014oua,Almeida:2015jua,deOliveira:2015xxd,Baldi:2016fql,Komiske:2016rsd,Kagan:2016wnu,Guest:2016iqz,Barnard:2016qma,Komiske:2017aww,ATLAS:2017dfg,Pearkes:2017hku,Kasieczka:2017nvn,Datta:2017rhs,Butter:2017cot,Datta:2017lxt,Egan:2017ojy,Schramm:2017frb,Louppe:2017ipp,Cheng:2017rdo,CMS:2017wtu,Komiske:2018oaa,Choi:2018dag,Macaluso:2018tck,Komiske:2018cqr,Kasieczka:2018lwf,Dreyer:2018nbf,Fraser:2018ieu,Lin:2018cin,Chen:2019uar,Datta:2019ndh,Qu:2019gqs,Chakraborty:2019imr,Lee:2019cad,Lee:2019ssx,Diefenbacher:2019ezd,Moreno:2019neq,Andreassen:2019txo,Moreno:2019bmu,Erdmann:2019blf,Li:2020grn,Bols:2020bkb,Chakraborty:2020yfc,Bernreuther:2020vhm,Lim:2020igi,Guo:2020vvt,Dolan:2020qkr,Mikuni:2020wpr,Li:2020bvf,Kagan:2020yrm,Erdmann:2020ovh,Dreyer:2020brq,Nakai:2020kuu,Bhattacharya:2020aid,CMS:2020poo,Andrews:2021ejw,Filipek:2021qbe,Mikuni:2021pou,Konar:2021zdg,Shimmin:2021pkm,Dreyer:2021hhr,Aguilar-Saavedra:2021rjk,Khosa:2021cyk,Gong:2022lye,Kim:2021gtv,Qu:2022mxj,ATL-PHYS-PUB-2017-003,ATL-PHYS-PUB-2020-014}, but relatively less work has been done on event-level classification~\cite{Louppe:2017ipp,Nguyen:2018ugw,Andrews:2018nwy,Lin:2018cin,Du:2019civ,Diefenbacher:2019ezd,Chung:2020ysf,Guo:2020vvt,Tannenwald:2020mhq,Ngairangbam:2020ksz}. Our work illustrates the potential benefits of full event-level classification. 

For simplicity, we will not consider SM backgrounds in this work; of course, these backgrounds are highly dependent on the Higgs final state. In certain decay modes such as $H \to ZZ^*\to4 \ell$~\cite{ATLAS:2020rej,ATLAS:2020wny,CMS:2021ugl,CMS:2021nnc}, the non-Higgs background is highly suppressed, so our work could directly apply there. For other decay modes where the SM background is less suppressed (e.g., $H\to\gamma\gamma$), we imagine the ``universal" VBF vs. GGF classifier could be combined with a Higgs decay classifier for full event classification including non-Higgs background rejection if necessary~\cite{ATLAS:2020pvn,ATLAS:2020qdt,CMS:2020xrn}.

An outline of our paper is as follows. In Sec.~\ref{sec:Sample Preparation}, we describe the simulation of our sample as well as the VBF pre-selection criteria and the numbers of training, validation, and testing sets for the classifier. In Sec.~\ref{sec:Classifier models}, we describe the classifiers used in this study. We show the results in Sec.~\ref{sec:Results}, which is comprised of a comparison of tagger performances, a discussion about what the event-level CNN has learned, and possible improvements of the BDT from adding information beyond the leading two jets.  In Sec.~\ref{sec: PTbalance}, we examine the $p_T$-balance of the full event and explore the possibility of the Higgs-decay-mode-agnostic classifier. Finally, we conclude in Sec.~\ref{sec:Conclusions}.  Appendix~\ref{sec:Architectures of various neural networks} lists the structures of all the classifier models considered in this study. Appendix~\ref{sec:Multi-stream CNN} examines an extension of CNN for our classification problem motivated by \cite{Chung:2020ysf}  and finds no further improvement.

\section{Sample Preparation \label{sec:Sample Preparation}}

We use \textsc{Madgraph5\_aMC@NLO} 2.7.3~\cite{Alwall:2014hca} (MG5) with parton distribution functions (PDFs) of \textsc{CT10}~\cite{Lai:2010vv} to generate Higgs plus up to three jets events starting from $pp$ collisions at $\sqrt{s} = 14$~TeV. The additional jets are matched using the MLM matching scheme with parameters $xqcut=30~\text{GeV}$ and $qcut=45~\text{GeV}$. For VBF we just use tree-level MG5, while for GGF we use a model generated by \textsc{FeynRules} 2.3.33~\cite{Alloul:2013bka} following the effective vertex method.

The samples are then showered and hadronized by \textsc{Pythia 8.245}~\cite{Sjostrand:2014zea,2006}, and finally passed through the \textsc{Delphes 3.4.2}~\cite{deFavereau:2013fsa} fast detector simulation. Note that for the showering of VBF we toggle on the local dipole recoil option~\cite{Cabouat:2017rzi} in \textsc{Pythia}, which models the emission of additional jets in a way compatible with QCD~\cite{Hoche:2021mkv,Jager:2020hkz,Konar:2022bgc}. This is achieved by specifying \texttt{SpaceShower:pTmaxMatch=2}, \texttt{SpaceShower:dipoleRecoil=on}, \texttt{TimeShower:QEDshowerByL=off}, and \texttt{PDF:lepton=off}. Also note that the detector configuration in \textsc{Delphes} is based upon the default ATLAS card, while the inputs of the jet cluster module are EFlow objects instead of the default Tower objects. The jet clustering is done by \textsc{FastJet} 3.3.2~\cite{Cacciari:2011ma} using the anti-$k_T$~\cite{Cacciari:2008gp} algorithm with $R=0.4$. Jets are required to have $p_T>25$~GeV in the simulation stage.

In our sample preparation, we let the Higgs decay to two photons and use their invariant mass cut to select the required Higgs production samples. Although we generate samples in this particular Higgs decay mode, as discussed in the Introduction, we will demonstrate later that the full-event classifiers trained on low-level inputs are actually agnostic to the Higgs decay products, in that their performance does not suffer when those decay products are removed. 

The samples used in the following analysis of this study are all extracted from the events passing the VBF pre-selection criteria, as inspired by experimental studies, that $N_{\gamma}\ge 2$, $120\le M_{\gamma\gamma}\le130$~GeV, $N_j\ge 2$, and $\Delta \eta_{jj}\ge 2$, with the jets in these criteria required to have $p_T>30$~GeV.
We have generated 500k events each for the VBF and GGF samples and, after the VBF pre-selection, are left with 177k events for VBF and 139k for GGF. 

Throughout this paper, we consider VBF as the signal and GGF as the background. For all of the event-level classifiers, the generated samples are split into training, validation, and testing sets as indicated in Table~\ref{tab:BDTevt_number}. Since for any event we take the leading two jets as the samples of the jet-level classifier (i.e., jet-CNN), the numbers of the samples in different sets of the jet-level classifier are twice as those in Table~\ref{tab:BDTevt_number}.

\begin{table}[htb]
    \centering
    \begin{tabular}{c c c c}
    \hline \hline
        & training & validation & testing \\
        \hline
        VBF events & 113k & 28k & 35k \\ 
        GGF events & 89k & 22k & 28k \\ 
        \hline \hline
    \end{tabular}
    \caption{Numbers of training, validation, and testing sets for event-level classifiers.}
    \label{tab:BDTevt_number}
\end{table}

\section{Classifier models\label{sec:Classifier models}}

\subsection{BDT}

\begin{table}[htb]
    \centering
    \begin{tabular}{lll}
    \hline \hline 
        Max depth && 3 \\ 
        Learning rate && 0.1 \\
        Objective && binary logistic \\
        Early stop && 10 epochs \\
        Evaluation metric && binary logistic \\
        \hline \hline
    \end{tabular}
    \caption{Hyperparameters of the BDT}
    \label{tab:BDTevt_hyper}
\end{table}

\begin{figure}[htb]
    \centering
    \includegraphics[scale=0.3]{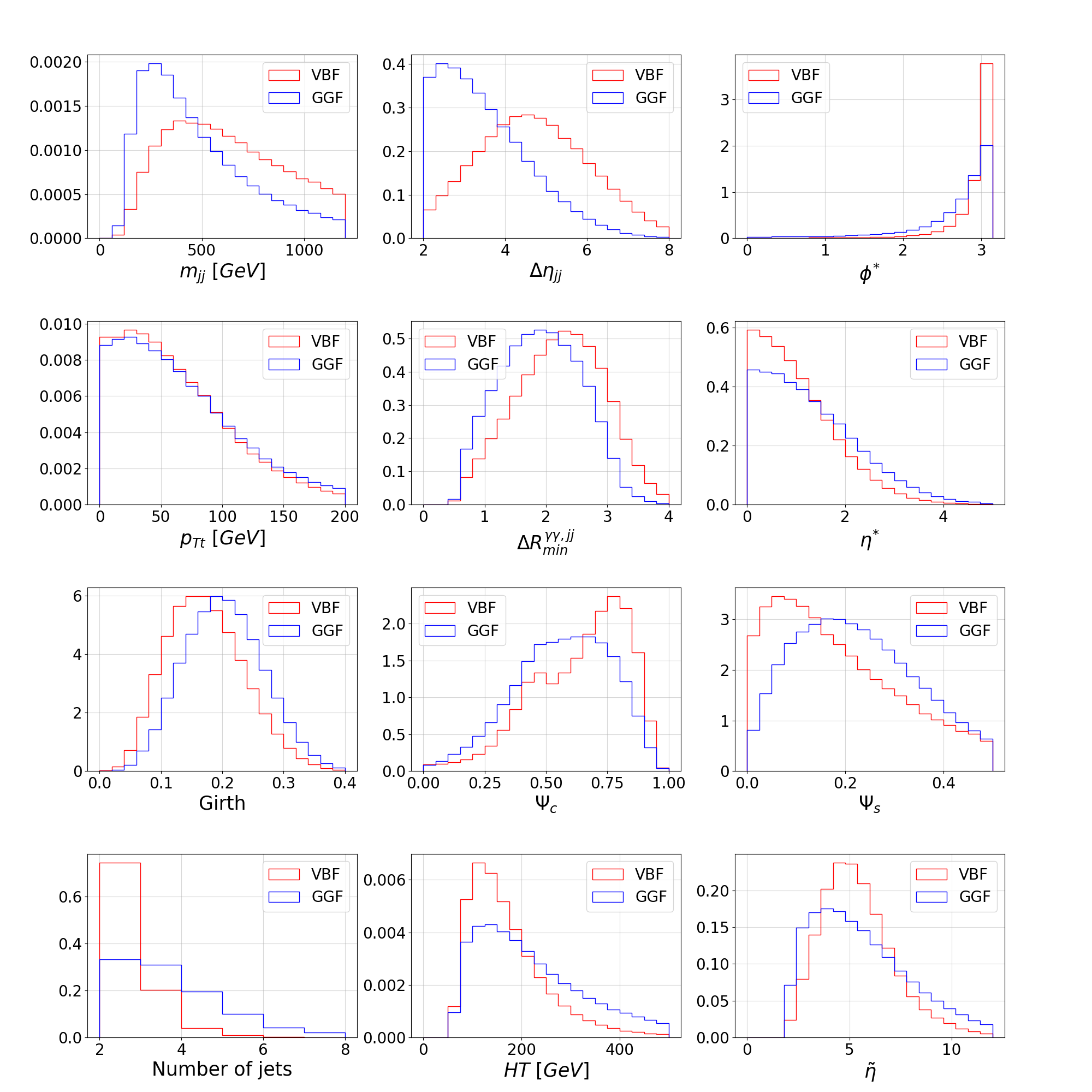}
    \captionsetup{justification=raggedright,singlelinecheck=false}
    \caption{Distributions of BDT input variables. All histograms are normalized so that the area under each curve is one.}
    \label{fig:input_features}
\end{figure}

We start by considering BDT models that are implemented in \textsc{XgBoost}~1.5.0~\cite{2016}. (The hyperparameters and the details of the BDT models are summarized in Table~\ref{tab:BDTevt_hyper}.) 
We train three different BDTs based on  the features summarized in Table~\ref{tab:BDT features} and Fig.~\ref{fig:input_features}.  The first, ``baseline", is based on six high level features from the study of VBF vs.\ GGF classification in Ref.~\cite{Chan:2017zpd}, which is inspired by ATLAS's setup~\cite{ATLAS:2018hxb}. This baseline BDT characterizes the discrimination power from the kinematics of the photons and the jets in the event.\footnote{We have checked that a simple DNN trained on these high-level features does not outperform the BDTs, so we will focus on BDTs as our baseline.}

Based on the experimental setup, Ref.~\cite{Chan:2017zpd} further considers the jet shape variables~\cite{Shelton:2013an} as additional input features, such as the girth summed over the two leading jets and the central/sided integrated jet shape. Including these jet shape variables leads to our second BDT, which we call ``baseline + shape".

Finally, we consider the benefits of replacing the human-engineered jet shape variables of ~\cite{Shelton:2013an,Chan:2017zpd} with the output of a jet-level CNN classifier trained on VBF vs GGF jets. We call this the ``baseline + jet-CNN" BDT. For more details on the jet-level CNN, see Section~\ref{sec:Jet-CNN}.

\begin{table}[htb]
    \centering
    \begin{tabular}{p{0.1\textwidth}p{0.85\textwidth}}
        \hline \hline
        \multirow{7}{*}{baseline} & 1.~$m_{jj}$, the invariant mass of $j_1$ and $j_2$ \\ 
        & 2.~$\Delta \eta_{jj}$, the absolute difference of the pseudo-rapidities of $j_1$ and $j_2$ \\
        & 3.~$\phi^*$, defined by the $\phi$-difference between the leading di-photon and di-jet \\
        & 4.~$p_{Tt}^{\gamma\gamma}$, defined by $\left|\left(\mathbf{p}_T^{\gamma_ 1}+\mathbf{p}_T^{\gamma_ 2}\right)\times\hat{t}\right|$, where $\hat{t}=\left(\mathbf{p}_T^{\gamma_ 1}-\mathbf{p}_T^{\gamma_ 2}\right)/\left|\mathbf{p}_T^{\gamma _1}-\mathbf{p}_T^{\gamma _2}\right|$\\
        & 5.~$\Delta R_{\gamma j}^{\text{min}}$, defined by the minimum $\eta$-$\phi$ separation between $\gamma_1$/$\gamma_2$ and $j_1$/$j_2$\\
        & 6.~$\eta^*$, defined by $\left|\eta_{\gamma_1\gamma_2}-\left(\eta_{j_1}+\eta_{j_2}\right)/2\right|$, where $\eta_{\gamma_1\gamma_2}$ is the pseudo-rapidity of the leading di-photon
        \\ \hline
        \multirow{3}{*}{shape} & 7.~the girth summed over the two leading jets $ \sum_{j=1}^2 g_j = \sum_{j=1}^2 \sum_{i\in J^j}^N \ p_{T,i}^j r_i^j / p_{T}^j $  \\
        & 8.~the central integrated jet shape $\Psi_c = \sum_{j=1}^2\sum_{i \in J^j}^N \ p_{T,i}^j (0<r_i^j<0.1) / (2p_{T}^j) $ \\
        & 9.~the sided integrated jet shape $\Psi_s = \sum_{j=1}^2\sum_{i \in J^j}^N \ p_{T,i}^j (0.1<r_i^j<0.2) / (2p_{T}^j)$ \\ \hline
        \multirow{2}{*}{jet-CNN} & 10.~the jet scores of the two leading jets, output by the jet-CNN, soon to be introduced in Section~\ref{sec:Jet-CNN} \\ 
        \hline \hline
    \end{tabular}
    \captionsetup{justification=raggedright,singlelinecheck=false}
    \caption{Summary of the features used in BDT. $j_1$ and $j_2$ mean respectively the $p_T$-leading and -subleading jets, while $\gamma_1$ and $\gamma_2$ mean respectively the $p_T$-leading and -subleading photons. 
    In the jet shape variables, $i$ represents the constituent of the jet and $r$ is the distance between the constituent and the jet axis.}
    \label{tab:BDT features}
\end{table}

\subsection{Jet-CNN \label{sec:Jet-CNN}}

In this subsection, we introduce the VBF vs.\ GGF jet-level CNN used in the ``baseline + jet-CNN scores" BDT described in the previous subsection.
The jet-level CNN is trained on jet images formed out of the leading two jets from the VBF and GGF events.\footnote{Another possible labeling scheme is to identify whether the jet is quark or gluon initiated, since VBF (GGF) events tend to contain more quark (gluon) jets. However, our trials show that both labeling schemes are equally useful when they are considered as features in the subsequent event-level BDT.  We will focus exclusively on the process-labeling in the following study.}  Our image pre-processing, which basically follows the procedure outlined in Ref.~\cite{Macaluso:2018tck}, contains image centralization, rotation, and flipping, followed by pixelation from the detector responses to the jet image. Finally, we pixelate the detector responses into images ($10\times10$ pixels) for each of the following four channels: Tower $E_T$, Tower hits, Track $E_T$ and Track hits. (Following the \textsc{Delphes} particle flow algorithm: ``Tower" means EFlowNeutralHadron or EFlowPhoton, and ``Track" means EFlowTrack.)

Our jet-CNN model starts from a Batch Normalization Layer~\cite{https://doi.org/10.48550/arxiv.1502.03167}, followed by several Convolution Layers and Average Pooling Layers, which capture the features of the images. The sizes of the filters in Convolution Layers and pools in Pooling Layers are all $2\times2$. Due to the relatively small size of the images ($10\times10$ pixels), the neural network (NN) does not need to be very deep.  Since the image size shrinks as it passes through a Pooling Layer, the number of Pooling Layers is restricted. After the Convolution and Pooling Layers, the images are then flattened and fully connected to three Dense Layers with 128 neurons respectively. The last Dense Layer with 2 neurons, activated by the SoftMax function, represents the final output score as probabilities.  All the other Dense Layers and Convolution Layers use the ReLU activation function~\cite{inproceedings}. The model structure is plotted in Fig.~\ref{fig:CNNjet_arch}. 

The CNNs in this study are all implemented in \textsc{TensorFlow} 2.0.0~\cite{tensorflow2015-whitepaper} with \textsc{Keras}~\cite{chollet2015keras} as its high-level API. We use Adam~\cite{https://doi.org/10.48550/arxiv.1412.6980} as our optimizer during the training stage with the categorical cross entropy loss function in all of our NN models. By monitoring the loss of the validation set, early stopping is implemented to prevent over-fitting in all of the NN and BDT models. The hyperparameters of the model are summarized in Table~\ref{tab:CNNjet_hyper}. 

\begin{table}[H]
    \centering
    \begin{tabular}{lll}
    \hline\hline
        Optimizer & & Adam \\ 
        Loss function & & categorical cross entropy \\
        Early stopping & & 20 epochs \\
        Batch size & & 1024 
        \\
    \hline\hline
    \end{tabular}
    \caption{Hyperparameters for the jet-CNN tagger.}
    \label{tab:CNNjet_hyper}
\end{table}

\begin{figure}[htb]
    \centering
    \includegraphics[scale=0.4]{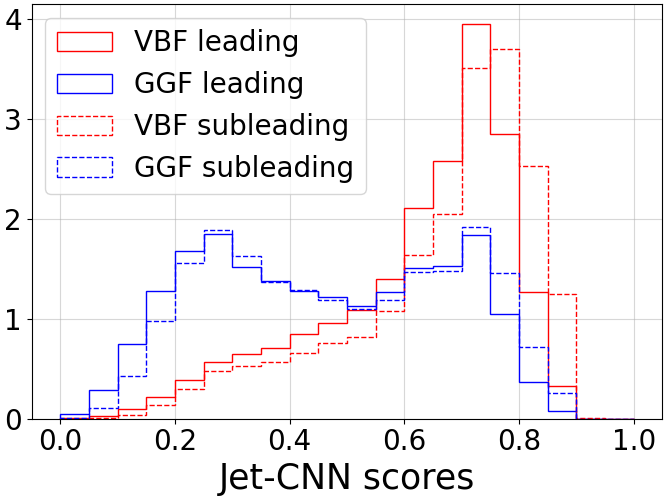}
    \hspace{0.5cm}
    \includegraphics[scale=0.41]{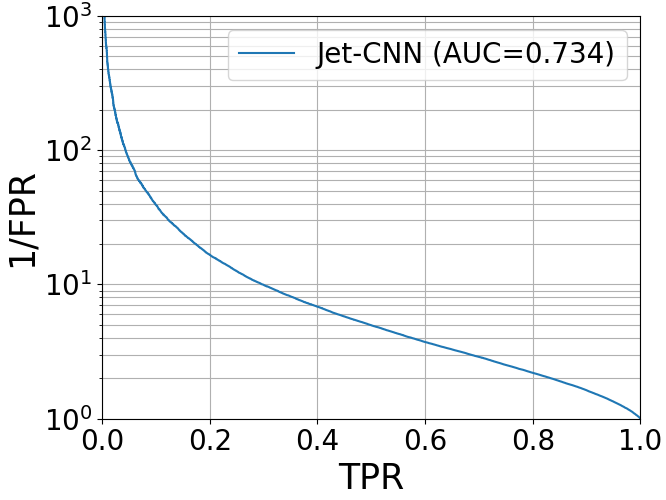}
    \captionsetup{justification=raggedright,singlelinecheck=false}
    \caption{Distributions of the jet-CNN scores (left) and the ROC curve of the jet-CNN (right). All histograms on the left are normalized so that each area under the curve is one.}
    \label{fig:input_jetCNN}
\end{figure}

Our jet-CNN takes a jet image as its input and outputs a score ranging from 0 (GGF-jet) to 1 (VBF-jet). The scores of leading and subleading jets can thus be useful features for subsequent event-by-event classification. The distributions of the jet-CNN scores and the ROC curve for the jet-CNN are shown in Fig.~\ref{fig:input_jetCNN}. The AUC of the jet-CNN is 0.734, which is less than an efficient classifier. However, we will show that the jet-CNN scores are indeed useful information in the subsequent event-level classification. Instead of training and testing separate taggers for the leading and subleading jets respectively, we utilize one tagger which is trained on mixed samples including the leading and subleading jets. Our trial shows that doing this way makes no loss of performance.

\subsection{Event-CNN \label{sec:Evt-CNN}}

A potentially more powerful way to perform event-level classification is to leverage the capabilities of deep learning to predict the VBF vs.\ GGF label directly from the lowest-level features of each event (in our case, the 4-vectors of all the particles in the event). In this paper we consider two approaches to this, a CNN trained on whole-event images, to be described in this subsection, and a self-attention model trained on sequences of the particle 4-vectors, to be described in the next subsection.

Our whole-event images are preprocessed similarly to the jet images of the previous subsection.  However, unlike jets, the whole event is not a localized object, nor is there an approximate boost or rotation invariance. So the preprocessing consists of just the following steps: we first move the $\phi$ coordinate of the weighted center to the origin, and flip the image vertically or horizontally to make the upper-right quadrant more energetic than all the other quadrants. Finally, the detector responses are pixelated into images with $40\times40$ pixels for each of the six channels, which includes the same four channels used in the jet-CNN and two additional ones recording the hits and $E_T$ of the isolated photons.

\begin{figure}[H]
    \centering
    \includegraphics[scale=0.3]{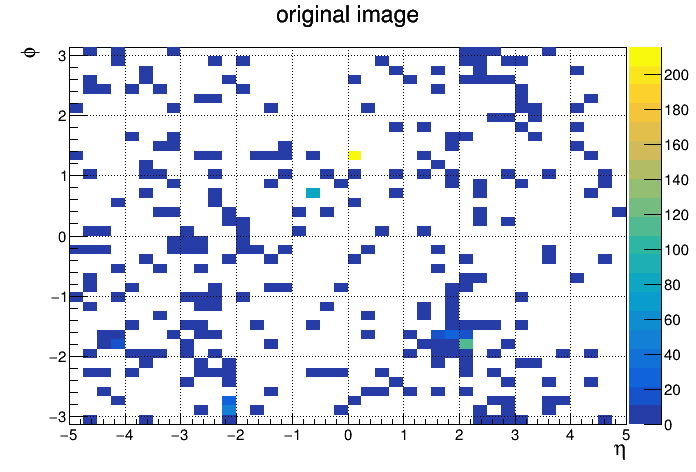}
    \hspace{0.5cm}
    \includegraphics[scale=0.3]{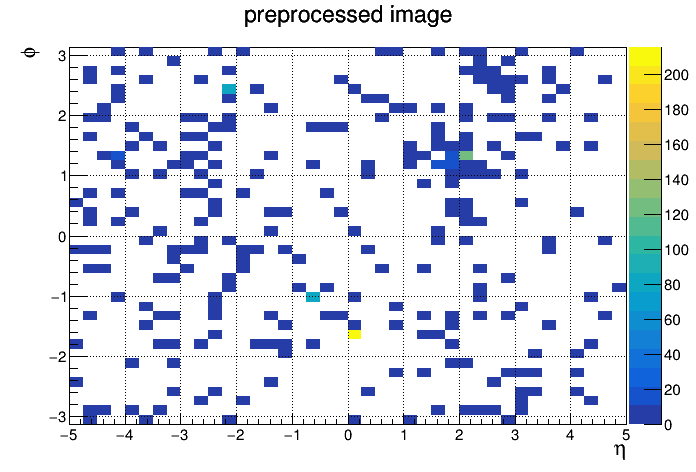}
    \captionsetup{justification=raggedright,singlelinecheck=false}
    \caption{The isolated photon $E_T$ and Tower $E_T$ combined with Track $p_T$ of an event without pre-processing (left) and after pre-processing (right). The color of each pixel indicates the energy in units of GeV.  }
    \label{fig:event_img}
\end{figure}

An example of single event images is shown in Fig.~\ref{fig:event_img}.  The left plot shows the isolated photon $E_T$ and Tower $E_T$ combined with Track $p_T$ of an event before the pre-processing, while the right plot is after the pre-processing.

We employ a toy ResNet model~\cite{he2015deep} in our event-CNN. Two Convolution Layers form a residual block in  ResNet. There are shortcuts connecting the residual blocks, enabling us to deepen our model without suffering from the degradation problem. The sizes of filters in the Convolution Layers and pools in the Pooling Layers are all $3\times3$. The detailed model structure of the event-CNN is shown in Fig.~\ref{fig:CNNevt_arch}. The hyperparameters are the same as those in Table~\ref{tab:CNNjet_hyper}.

In order to extract information from both the local jet-level and global event-level features, Ref.~\cite{Chung:2020ysf} adopts a two-stream CNN architecture, where one stream processes an image of the highest $p_T$ non-Higgs jet in the event, and the other stream processes the full-event image. Motivated by this, we further study the performance of an extension of our full-event CNN in Appendix~\ref{sec:Multi-stream CNN}, using a similar structure containing three streams of CNN, dealing with event images and leading two jet images respectively. However, we find no improvement from our original single-stream event-CNN. This does not contradict the works of Ref.~\cite{Chung:2020ysf} since they did not compare the performance of their two-stream CNN against a single-stream CNN consisting of just the full-event classifier.

\subsection{Self-attention}

For comparison, we also consider another whole-event low-level-feature classifier based on the technique of self-attention~\cite{DBLP:journals/corr/LinFSYXZB17}, which is used in the famous Transformer model~\cite{https://doi.org/10.48550/arxiv.1706.03762} dealing with sequence-to-sequence tasks. The original motivation of this model is to use the multi-head attention layers to capture the correlation among elements in the input sequence. Inspired by this idea, instead of representing an event as an image, we view the event as a sequence, where the elements of the sequence are the $p_T$, $\eta$, $\phi$, and electric charge of the 100 highest-$p_T$ reconstructed particles in the event  (with zero padding for events with fewer than 100 particles). In principle, the self-attention network could be advantageous over event-level images, because  it is not subject to the information loss induced by pixelation. Also, a nice property of the self-attention mechanism is that it preserves permutation invariance of the inputs (as does a CNN). 

The implementation of the self-attention model is based on \textsc{TensorFlow} 2.5.0 and \textsc{Keras}. The model structure of the self-attention model is shown in Fig.~\ref{fig:trans_arch}. There are three five-head attention layers at the beginning, followed by a Global Average Pooling (GAP) Layer, which converts the sequence of detector responses into a single vector by taking the element-wise average. Dense Layers are not implemented before the GAP Layer to keep permutation invariance of the input sequence. Then the model is passed into seven Dense Layers. The hyperparameters are listed in Table~\ref{tab:trans_hyper}.

\begin{table}[H]
    \centering
    \begin{tabular}{lll}
    \hline \hline
        Optimizer && Adam \\ 
        Loss function && categorical crossentropy \\
        Early stopping && 50 epochs \\
        Batch size && 1024 \\
        \hline \hline 
    \end{tabular}
    \caption{Hyperparameters of the self-attention model.}
    \label{tab:trans_hyper}
\end{table}

\section{Results\label{sec:Results}}

\subsection{Comparison of methods}

\begin{figure}[H]
    \centering
    \includegraphics[scale=0.25]{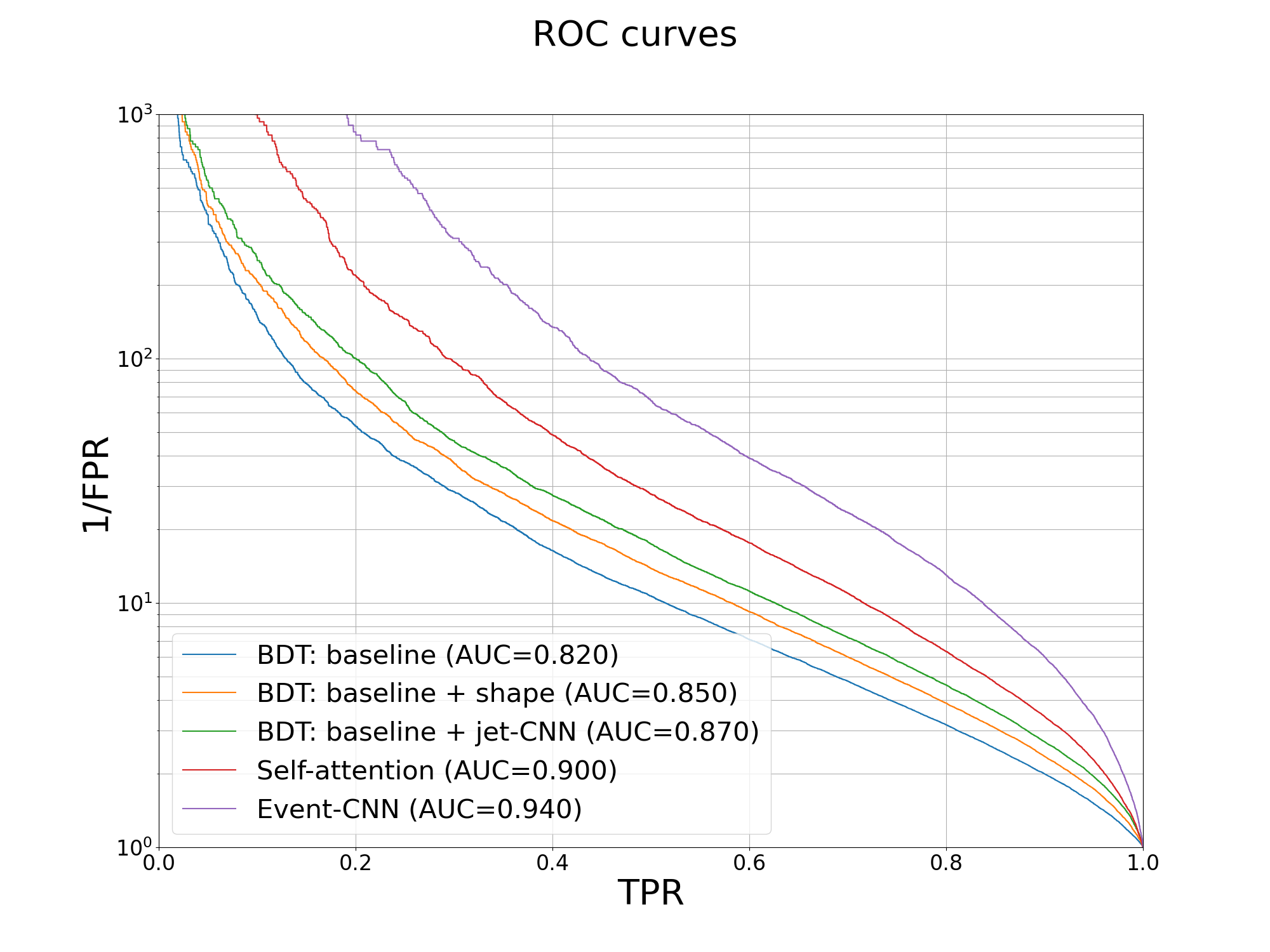}
    \caption{ROC curves of several event-level classifiers.}
    \label{fig:evt_ROC}
\end{figure}

\begin{table}[H]
    \centering
    \begin{tabular}{l c c}
        \hline \hline 
        & ~FPR~ & ~AUC~ \\ \hline 
        BDT: baseline & 0.035 & 0.820 \\ 
        BDT: baseline + shape & 0.027 & 0.850 \\ 
        BDT: baseline + jet-CNN & 0.022 & 0.870 \\ 
        Self-attention & 0.010 & 0.900 \\ 
        Event-CNN & 0.003 & 0.940 \\ 
        \hline \hline
    \end{tabular}
    \caption{Performance comparison at TPR = 0.3.}
    \label{tab:FPR_AUC}
\end{table}

The performance of the event-level classifiers defined in the previous section is shown in Fig.~\ref{fig:evt_ROC}.  As an explicit example, Table~\ref{tab:FPR_AUC} lists both false positive rates (FPRs) and AUCs at the working point where the true positive rate (TPR) is fixed at 0.3. From Fig.~\ref{fig:evt_ROC} and Table~\ref{tab:FPR_AUC}, we can easily compare the different event-level classifiers. 

First of all, ``BDT: baseline'' has the lowest AUC since it only considers the high-level kinematic features in an event.  Indeed, including additional information on the jet shape variables can improve a little, but not as much as using the jet-CNN score as an input.  Notably, our jet-CNN scores serve as a better feature than the jet shape variables, with the former reducing the FPR from the baseline by a factor of 1.6 while the latter only by a factor of 1.3.  Therefore, despite a low AUC of the jet-CNN as shown in Fig.~\ref{fig:input_jetCNN}, its score still provides valuable information. We have also checked that combining jet shape variables and jet-CNN scores in the input features together did not provide extra improvement in the AUC, indicating that the jet-CNN has learned all the information contained in the human-engineered jet shape variables. 

Second, we see that our self-attention model and event-CNN both perform better than the BDTs.  This is understandable because the BDTs only take into account high-level variables or features of the two leading jets and photons only, while the self-attention and event-CNN taggers take in the entire event and catch more features therein. 

Finally, the event-CNN is the most powerful classifier among all considered taggers.  Its inverse FPR is roughly a factor of 2.5 better than the self-attention model for most of the TPR.  Its AUC reaches 0.940 and the FPR is reduced by a factor of 11.7 from the baseline at the assumed working point in Table~\ref{tab:FPR_AUC}. 

\subsection{Saliency maps of event-CNN}

\begin{figure}[htb]
    \centering
    \includegraphics[scale=0.38]{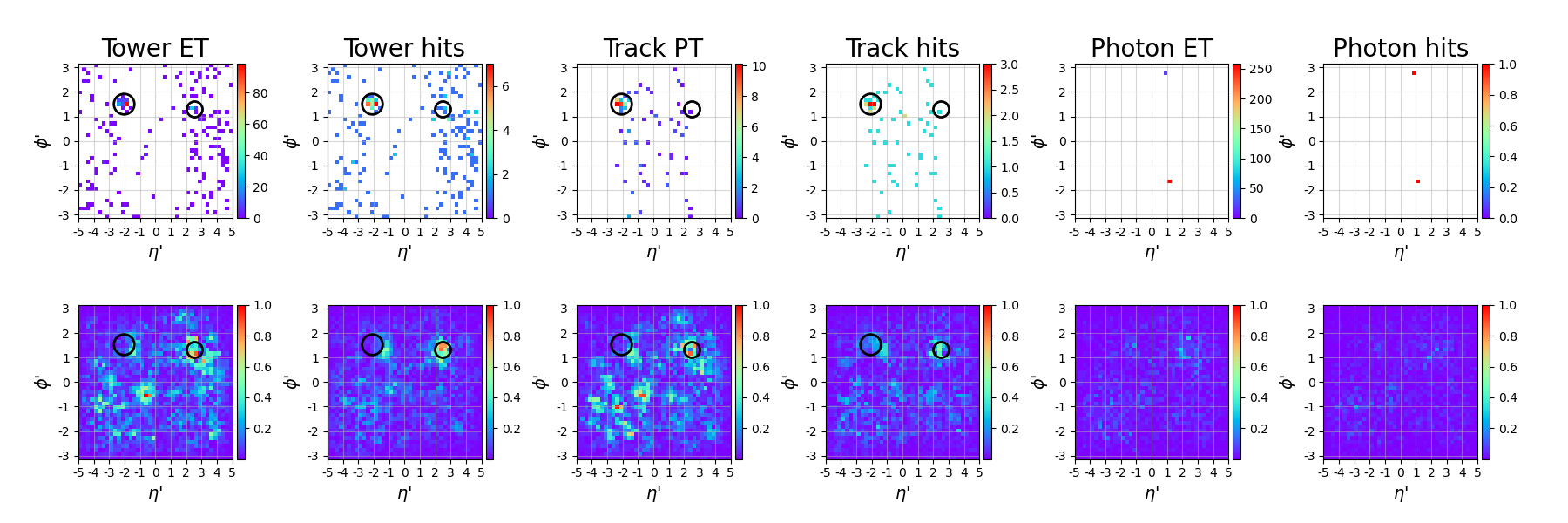}
    \captionsetup{justification=raggedright,singlelinecheck=false}
    \caption{A VBF event with a high event-CNN score.  The upper six plots show the raw inputs of the model, while the lower counterparts are the saliency maps calculated by the corresponding normalized channels.  The black circles show the locations of the clustered jets, with the circle size indicating the ordering in $p_T$. The color maps of the upper row indicate the actual input. The unit is GeV for Tower $E_T$, Track $p_T$, and isolated photon $E_T$, and counts for Tower hits, Track hits, and isolated photon hits. The color maps of the lower row indicate the relative saliency. }
    \label{fig:samap_VBF1}
\end{figure}

\begin{figure}[htb]
    \centering
    \includegraphics[scale=0.38]{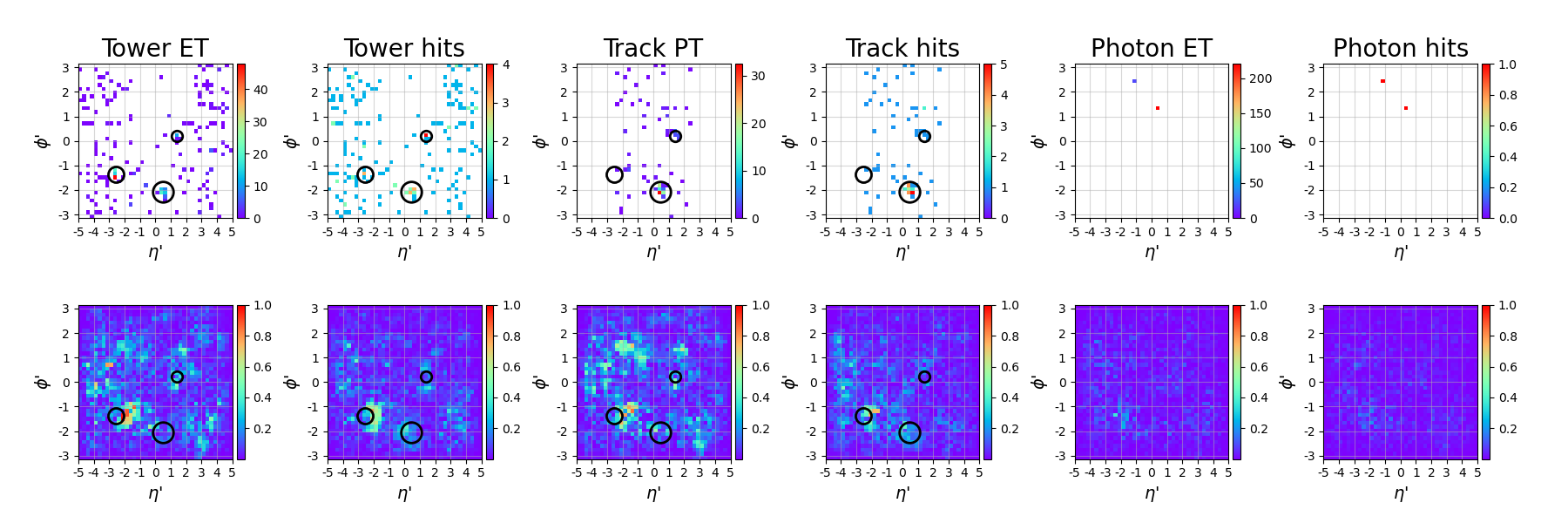}
    \caption{Same as Fig.~\ref{fig:samap_VBF1}, but for a GGF event with a low event-CNN score.}
    \label{fig:samap_GGF1}
\end{figure}

To further investigate what the event-CNN has learned, we examine its saliency maps~\cite{simonyan2014deep}.  Let the input pixel $x$ be identified as $x_{c,h,w}$, where $c$ is the channel index, $h$ is the height index, and $w$ is the width index. The saliency is defined by the gradient of the $i$-th class score $P^i$ with respect to the input pixel $x_{c,h,w}$, 
\begin{equation}
    w^i_{c,h,w} \equiv \frac{\partial P^i}{\partial x_{c,h,w}} ~,
\end{equation}
where the gradient is calculated by back-propagation. In our case, we only deal with binary classifiers, so it suffices to only consider the VBF class score $P$. Putting $w_{c,h,w}$ together according to the indices, one can obtain the saliency maps. However, what we are actually interested in is the saliency according to the standardized pixels $y_{c,h,w}$ which have no scale difference across channels, 
\begin{equation}
    x_{c,h,w}\to y_{c,h,w}=\frac{x_{c,h,w}-\mu_c}{\sigma_c} ~,
\end{equation}
where ${\sigma_c}^2$ and $\mu_c$ are the variance and mean of the channel $c$ in the whole sample, including the training, validation, and testing sets. Hence, we will consider the following gradient, 
\begin{equation}
    \tilde{w}_{c,h,w} \equiv \frac{\partial P}{\partial y_{c,h,w}} = 
     w_{c,h,w}\times \sigma_c ~.
\end{equation}
Finally, we arrange $\tilde{w}$ according the $c,h,w$ indices and then plot its absolute value $|\tilde{w}_{c,h,w}|$ to get the saliency maps as the lower row in Fig.~\ref{fig:samap_VBF1} and \ref{fig:samap_GGF1}.

We utilize the visualization toolkit \textsc{tf-keras-vis}~0.8.0~\cite{Kubota_tf-keras-vis_2021} to implement the saliency maps of our event-CNN tagger. In the following, we pick as examples a VBF event (Fig.~\ref{fig:samap_VBF1}) with a high CNN score (i.e., more VBF-like) and a GGF event (Fig.~\ref{fig:samap_GGF1}) with a low CNN score (i.e., more GGF-like). In the plots, the clustered jets are marked by black circles, with their sizes indicating the jet's ordering in $p_T$. The color maps of the upper row indicate the actual value of the input, with the unit being GeV for Tower $E_T$, Track $p_T$, and isolated photon $E_T$ and counts for Tower hits, Track hits, and isolated photon hits. In contrast, the color maps of the lower row indicate the relative saliency, i.e.\ the most salient pixel is scaled to one in plotting,
\begin{equation}
    \left|\tilde{w}_{c,h,w}\right|\to\frac{\left|\tilde{w}_{c,h,w}\right|}{
    \displaystyle \max_{c,h,w}\left\{\left|\tilde{w}_{c,h,w}\right|\right\}} ~.
\end{equation}

From the saliency maps, we observe that the CNN model generally focuses on the locations with more hadronic activities, as anticipated, because the jets contain crucial information for the classification of VBF and GGF events.  In addition, the CNN is also seen to make use of lower $p_T$ jets and hadronic activity that falls below the jet $p_T$ threshold (set to 30~GeV in this work). This explains why the event-CNN performs better than the BDT. In our setup of the BDTs, we do not feed the information of the third jet into the model. Moreover, the input of the BDTs relies on our knowledge about what kind of high-level features is beneficial and hence cannot make use of unclustered energy in an event. Finally, we can observe that the event-CNN is much more focused on where jets are than the locations of photons, which indicates that the photon information is not crucial in the classification. This sheds light on the possibility of the Higgs-decay-mode-agnostic classifier which solely relies on the jet information. Details will be described in Section~\ref{sec: PTbalance}.

\subsection{Improvements of BDTs \label{sec:Improvements of BDTs}}

In this subsection, we investigate more about how the BDTs, which rely on high-level kinematic variables as the features for training, can be further improved.   Based on the study of the saliency maps in the previous subsection, we are motivated to consider information about additional hadronic activity in the event beyond the leading two jets. So we will study the benefits of including the 4-vector momentum of the third hardest jet, as well as inclusive kinematic variables that take all jets into account.

\begin{itemize}
    \item 4-vector momentum of the third jet in $p_T$ ordering, which is denoted as ``j3vec;'' 
    \item $HT=\sum\limits_{j\in \text{jets}}p_T^j$, which characterizes the $p_T$ distribution of the jets;
    \item $\tilde{\eta} = \sum\limits_{j\in \text{jets}}\left|\eta^j\right|$, which characterizes the positional distribution of the jets; and
    \item number of jets.
\end{itemize}

We will call the set of features including $HT$, $\tilde{\eta}$, and the number of jets as a ``jet-profile.'' The normalized distributions of $HT$, $\tilde{\eta}$, and the number of jets are already shown in the last row of Fig.~\ref{fig:input_features}.

\begin{figure}[htb]
    \centering
    \includegraphics[scale=0.25]{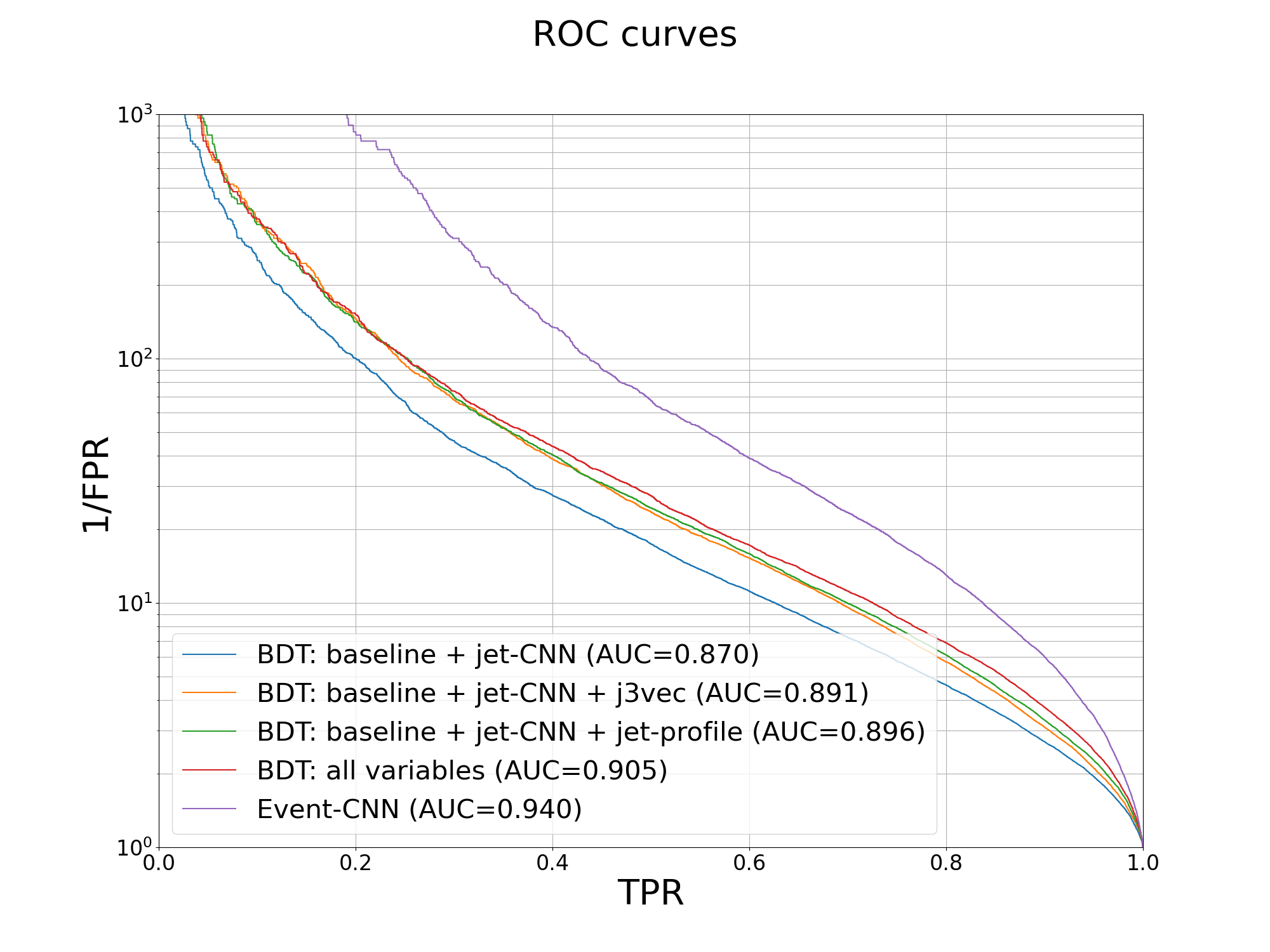}
    \caption{ROC curves of BDT trained on additional high-level features.}
    \label{fig:comp_kin1}
\end{figure}

We are interested in how the additional information improves the best BDT we have so far, so we will add these extra variables to ``BDT: baseline + jet-CNN." The ROC curves of the BDTs trained with further inputs of these additional variables are plotted in Fig.~\ref{fig:comp_kin1}.  From the AUCs, we can see both the additional 4-vector momentum of the third jet and the jet-profile can improve the performance of classification.  Their improvements are comparable to each other, as seen from the ROC curves as well as the similar AUCs. We have also checked that adding the additional 4-vector momentum and the jet-profile together into the BDT does not further improve the AUC, which is a piece of direct evidence that these two sets of variables provide equivalent information to the BDTs. The reason is that the crucial information contained in both sets is the existence of the third jet. A characteristic of GGF events is that they tend to have more than two jets, which can be seen in the distribution of the number of jets in Fig.~\ref{fig:input_features}. By examining the actual trees in the BDTs trained by the 4-vector momentum of the third jet and the jet-profile, respectively, we indeed find that the existence of the third jet provides a clear separation between VBF and GGF events and therefore plays an important role in both cases.

Finally, in ``BDT: all variables'',  we consider all the high-level features, including event-related characteristics (i.e., $m_{jj}$, $\Delta \eta_{jj}$, $\phi^*$, $p_{Tt}^{
\gamma\gamma}$, $\Delta R_{\gamma j}^{\text{min}}$, $\eta^*$, $HT$, $\tilde{\eta}$, and the number of jets), and jet-related information of the three leading jets (i.e., 4-vector momenta, jet-CNN scores, and the girth, central/sided integrated jet shape of each jet without taking summation or average). This BDT achieves the best AUC, 0.905, among all the other BDTs and improve the baseline significantly. However, despite this sizable improvement from the baseline, the event-CNN still outperforms ``BDT: all variables'' with an even larger AUC.

\section{A Higgs-decay-agnostic VBF vs.\ GGF classifier \label{sec: PTbalance}}

In the Introduction, we noted that event-level classifiers trained on low-level inputs could potentially be agnostic to the Higgs decay mode, due to the scalar nature of the Higgs and the $p_T$-balance of the whole event. Here we will explore this idea further, by seeing to what extent $p_T$ balance allows the Higgs momentum to be predicted from the hadronic activity in the event, and then to what extent our classifiers suffer when the Higgs decay products are removed from the event.

\begin{figure}[H]
    \centering
    \includegraphics[scale=0.6]{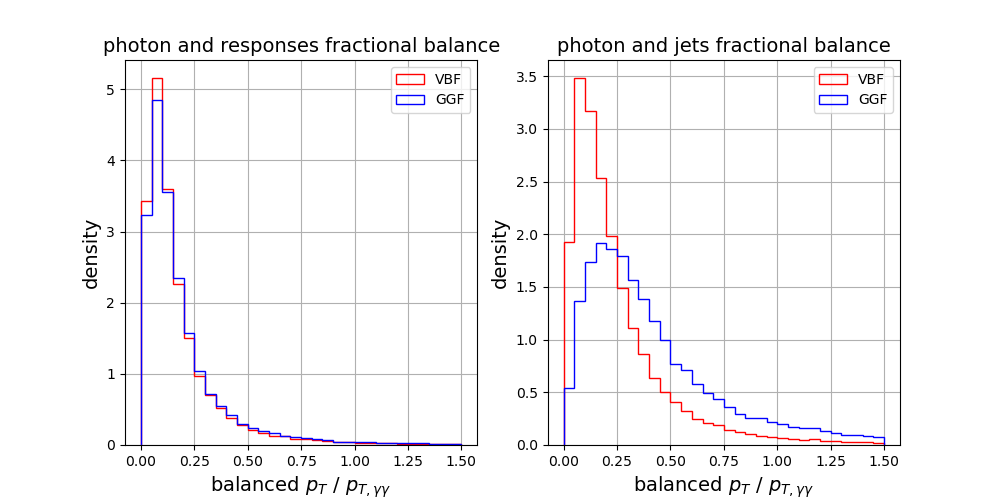}
    \captionsetup{justification=raggedright,singlelinecheck=false}
    \caption{The fractional $p_T$-balance of the leading di-photon and other objects, non-photon responses on the left and up to leading three jets on the right. To calculate the balance, we first vector-sum the momenta of the di-photon and other objects, and then take its transverse momentum. Finally, the balanced $p_T$ is divided by the $p_T$ of the di-photon.}
    \label{fig:PTbalance4.1}
\end{figure}

Shown in the left plot of Fig.~\ref{fig:PTbalance4.1} are histograms of the $p_T$ balance of the whole event, $|\sum_{i\in {\rm reconstructed\,\,\,particles}} \vec p_{Ti}|$, normalized by the $p_T$ of the Higgs. We see that the $p_T$ is well-balanced amongst the low-level features, so the Higgs transverse momentum can be well-reconstructed from the non-photon reconstructed particles. Meanwhile, the right plot of Fig.~\ref{fig:PTbalance4.1} depicts the $p_T$ balance between the photons and the leading three jets, again normalized by the $p_T$ of the Higgs. Here we see that while the leading three jets can capture the $p_T$ information of the photons to some extent, it is not as informative as the responses and, therefore, the balance is not as complete.

\begin{figure}[H]
    \centering
    \includegraphics[scale=0.25]{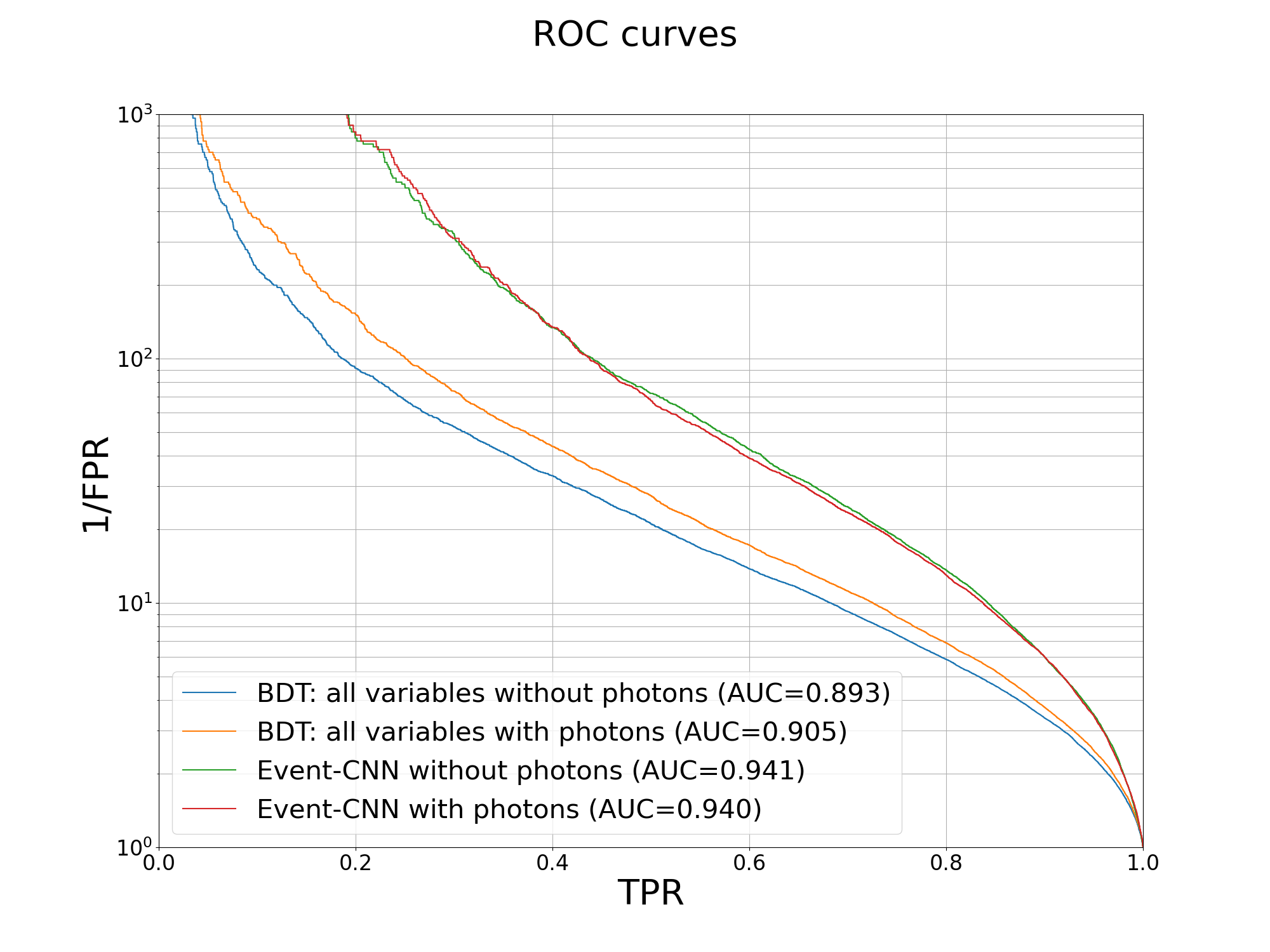}
    \captionsetup{justification=raggedright,singlelinecheck=false}
    \caption{ROC curves of event-level classifiers with and without the photon information.}
    \label{fig:evt_ROC4}
\end{figure}

Fig.~\ref{fig:evt_ROC4} shows the impact of removing the Higgs decay products (in our case, the two photons) from the event before training the VBF vs.\ GGF classifiers. We see that removing photon information from the event-CNN hardly changes the AUC. On the other hand, removing photon information from the high-level features for the BDT reduces the AUC from 0.905 to 0.893.\footnote{In more detail, in Fig.~\ref{fig:evt_ROC4}, ``all variables with photons'' refers to the feature set used in ``BDT: all variables'' in Fig.~\ref{fig:comp_kin1}, while ``all variables without photons'' refers to the same feature set  but with the photon-related variables (i.e., $\phi^*$, $p_{Tt}^{\gamma\gamma}$, $\Delta R_{\gamma j}^{\text{min}}$, $\eta^*$) all excluded.} The degradation in performance in the BDT is still not very large, but it is larger than that of the event-CNN. This is completely in line with the histograms shown in Fig.~\ref{fig:PTbalance4.1}. 

All in all, we confirm here that due to the $p_T$ balance of the events, the performance of VBF vs.\ GGF classification does not depend much on the Higgs decay products, especially for the whole-event CNN that is based on low-level inputs. This raises the intriguing possibility that one could train a single VBF vs.\ GGF classifier that is agnostic to the Higgs decay mode, and could be applied equally optimally to a variety of Higgs decay channels in a uniform way. This could have benefits for data-driven calibration and reducing systematic uncertainties associated with VBF tagging.

\section{Conclusions\label{sec:Conclusions}}

In this paper, we have studied machine learning approaches to event-level classification of Higgs production modes, focusing on the important problem of VBF vs.\ GGF discrimination. Building on previous studies~\cite{Chan:2017zpd,Chung:2020ysf}, we have shown that full-event deep learning classifiers that utilize low-level inputs (full-event images, sequences of particle 4-vectors) significantly outperform classifiers based on high-level physics features (kinematics, jet shapes). We have explored both CNNs trained on full-event images, and permutation-invariant self-attention networks trained on sequences of particle 4-vectors. Although the full-event CNN achieved the best performance in our studies -- improving beyond the baseline shallow network by more than a factor of $4-15$ in background rejection across a wide range of signal efficiencies -- perhaps our work provides a useful starting point for further optimization of the attention-based approach.

We also studied why the event-level CNNs perform so much better than the shallow networks based on high-level features. Using saliency maps we saw how additional jets in the event beyond the first two contribute to the CNN classification, as well as unclustered hadronic activity (jets below the $p_T$ threshold). By adding high-level features derived from these additional jets, we have confirmed that the performance of the shallow networks can indeed be improved and be brought somewhat closer to the event-level CNN.

Finally, in this work we have gone beyond previous approaches and explored the possibility of a VBF vs.\ GGF classifier that is agnostic to the Higgs decay mode. A classifier trained on the low-level information of the full event should be able to reconstruct the Higgs transverse momentum from $p_T$ balance and, since the Higgs is a scalar,  its decay products (at least when it decays to electroweak states) should be well-factorized from the rest of the event. Therefore, a full-event, low-level classifier should be largely independent of the Higgs decay channel. We have taken the first steps towards verifying that in this work, by showing how the performance of the full-event CNN is virtually unchanged when trained on the events with and without the diphotons from the Higgs decay.

Some future directions of our work include: generalizing our work to more Higgs production modes (e.g., $ZH$, $WH$ and $ttH$) using a multi-class classifier; further fleshing out our idea of a decay-agnostic classifier by studying other Higgs decay modes besides $H\to \gamma\gamma$; studying even more recent deep learning classifiers such as graph networks; adding particle interaction information into the event-level self-attention model (as was inspired by Ref.~\cite{Qu:2022mxj}); and incorporating symmetries such as Lorentz invariance into the architecture of the neural network to achieve even better performance (as was done recently for top-tagging in Ref.~\cite{Gong:2022lye}).

\acknowledgments
We are grateful to Tae Min Hong and Yi-Lun Chung for helpful discussions.  We also thank Kai-Feng Chen and Iftah Galon for their participation in this project at the early stage. In addition, we thank Alexander Karlberg for pointing out the importance of the local dipole recoil scheme for VBF. CC and SW were supported in part by the Ministry of Science and Technology of Taiwan under Grant Nos.~MOST-108-2112-M-002-005-MY3 and 111-2112-M-002-018-MY3. The work of DS was supported by DOE grant DOE-SC0010008.

\appendix

\section{Architectures of various neural networks
\label{sec:Architectures of various neural networks}}

\begin{figure}[H]
    \centering
    \includegraphics[scale=0.3]{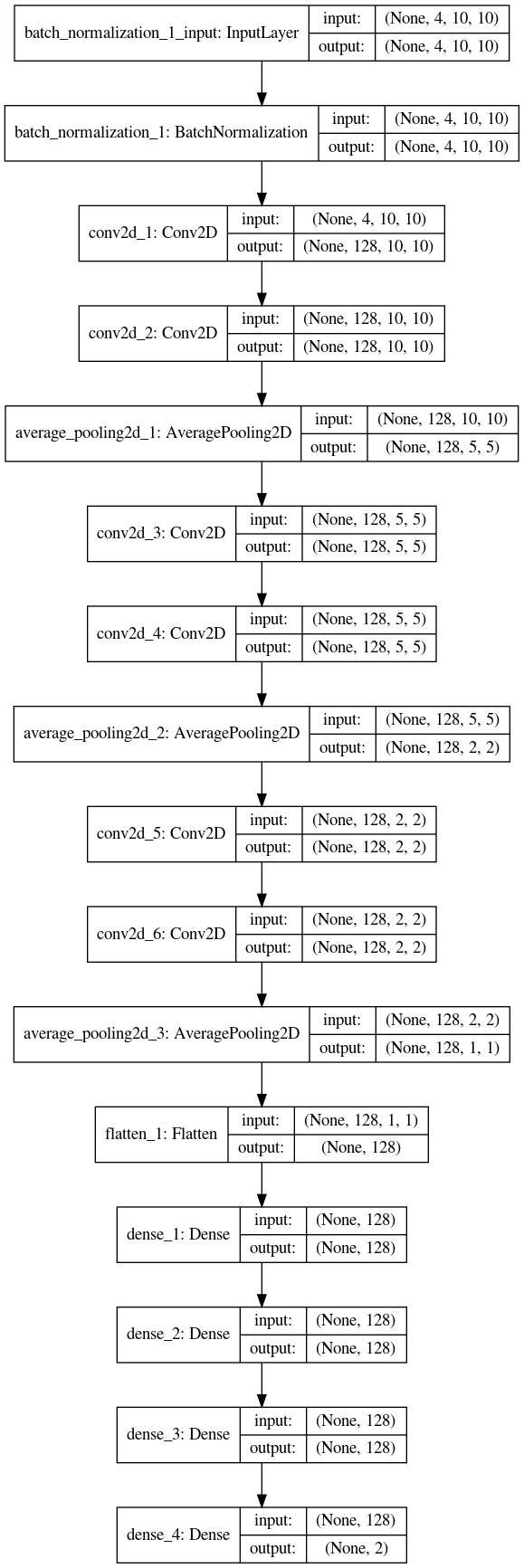}
    \caption{The model structure of the jet-CNN.}
    \label{fig:CNNjet_arch}
\end{figure}

\begin{figure}[H]
    \centering
    \includegraphics[scale=0.3]{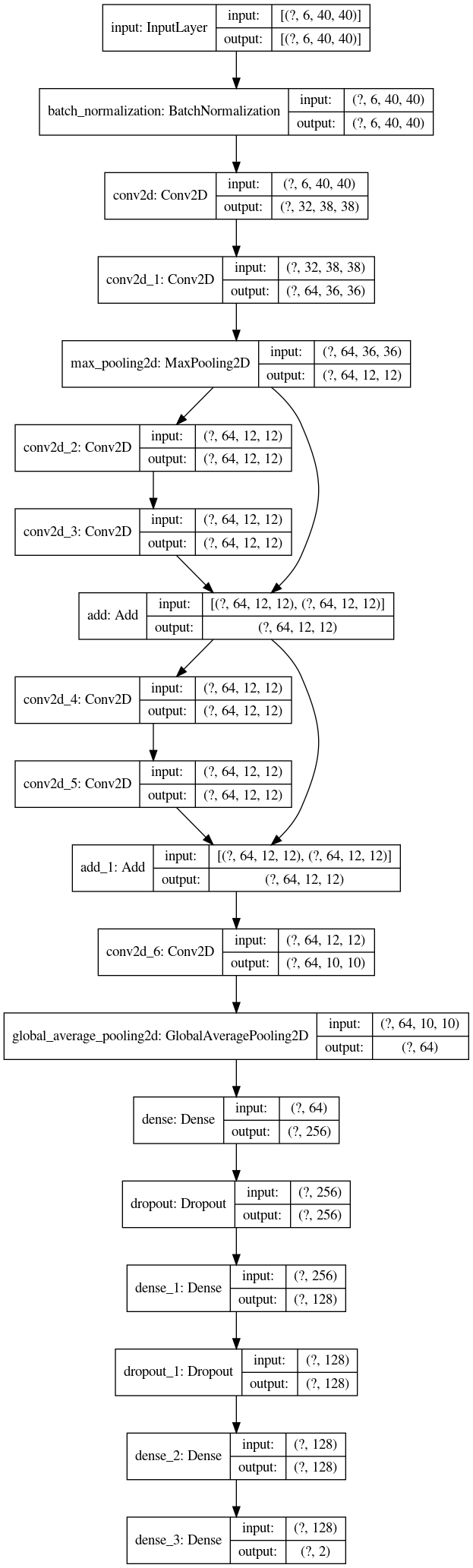}
    \caption{The model structure of the event-CNN.}
    \label{fig:CNNevt_arch}
\end{figure}

\begin{figure}[H]
    \centering
    \includegraphics[scale=0.3]{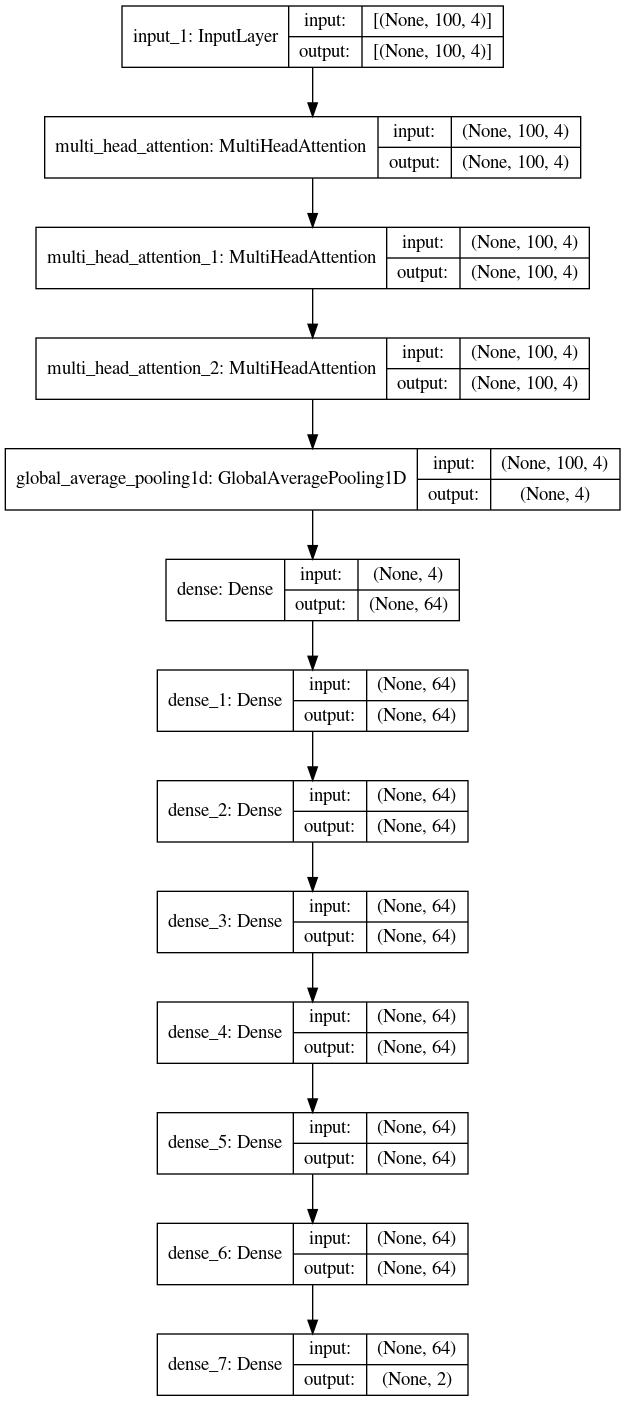}
    \caption{The model structure of the self-attention model.}
    \label{fig:trans_arch}
\end{figure}

\begin{figure}[H]
    \centering
    \includegraphics[scale=0.25]{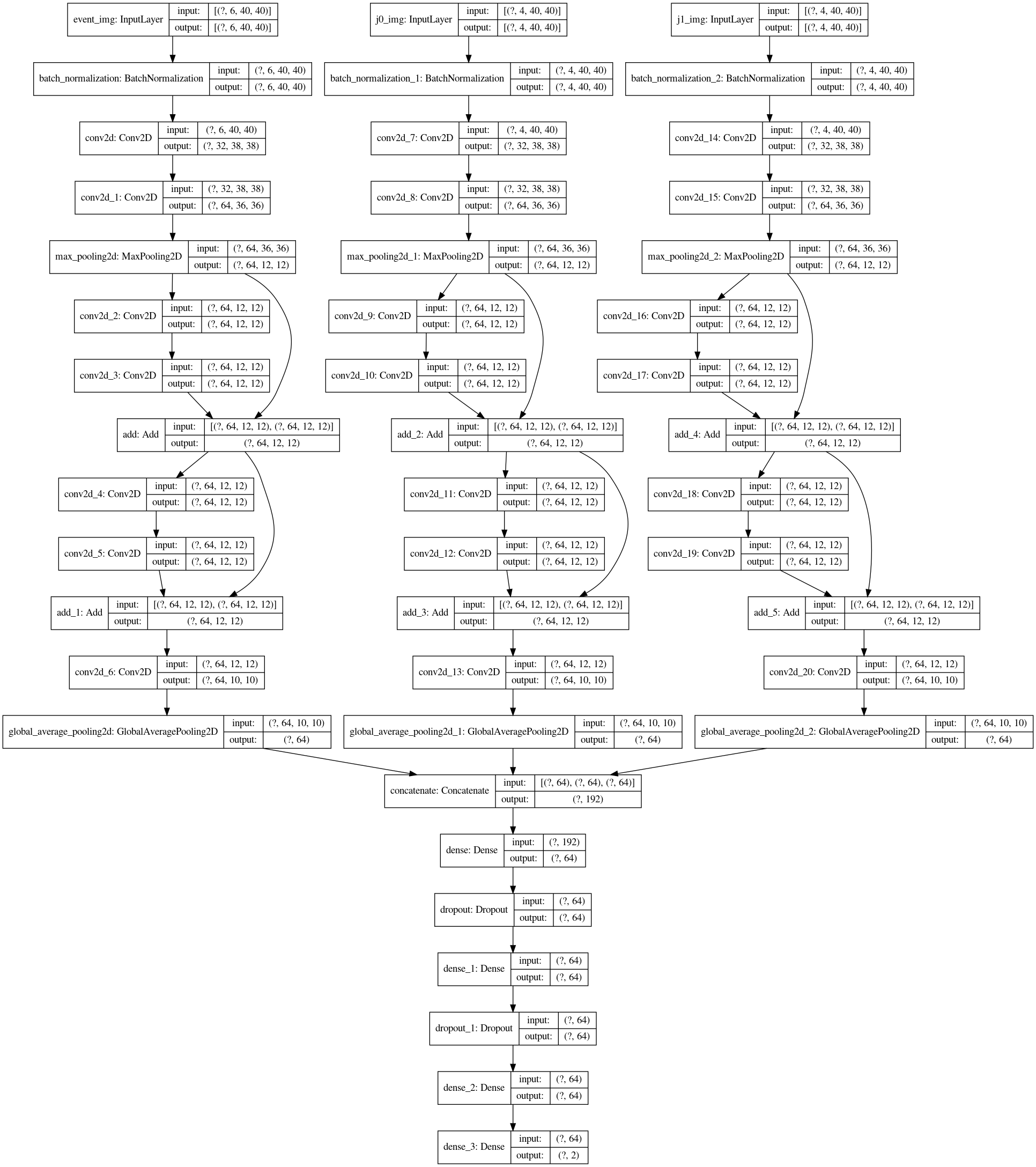}
    \caption{The model structure of the multi-stream CNN model.}
    \label{fig:3CNN_arch}
\end{figure}

\section{Multi-stream CNN
\label{sec:Multi-stream CNN}}

In this section, we examine an extension of CNN. In Ref.~\cite{Chung:2020ysf}, an architecture called 2CNN extracts event-level and jet-level features simultaneously with two streams. One stream applies filters on event images, and the other deals with the leading non-Higgs jet images. Then the two streams are connected together and combined to form a single model. Inspired by this study, we want to investigate the possible improvement using this multi-stream architecture. 

We adopt a three-stream CNN where one stream applies filters on the event images, and the other two process the images of the two leading jets respectively. Each stream is a toy ResNet model used in Sec.~\ref{sec:Evt-CNN}. We will call this architecture ``event + 2jet-CNN.'' The model structure is shown in Fig.~\ref{fig:3CNN_arch}. All of the event images and jet images are pixelated into $40\times40$ pixels. The images are pre-processed as in Sec.~\ref{sec:Jet-CNN} and ~\ref{sec:Evt-CNN}. The hyperparameters are the same as those in Table~\ref{tab:CNNjet_hyper}. 

\begin{figure}[H]
    \centering
    \includegraphics[scale=0.25]{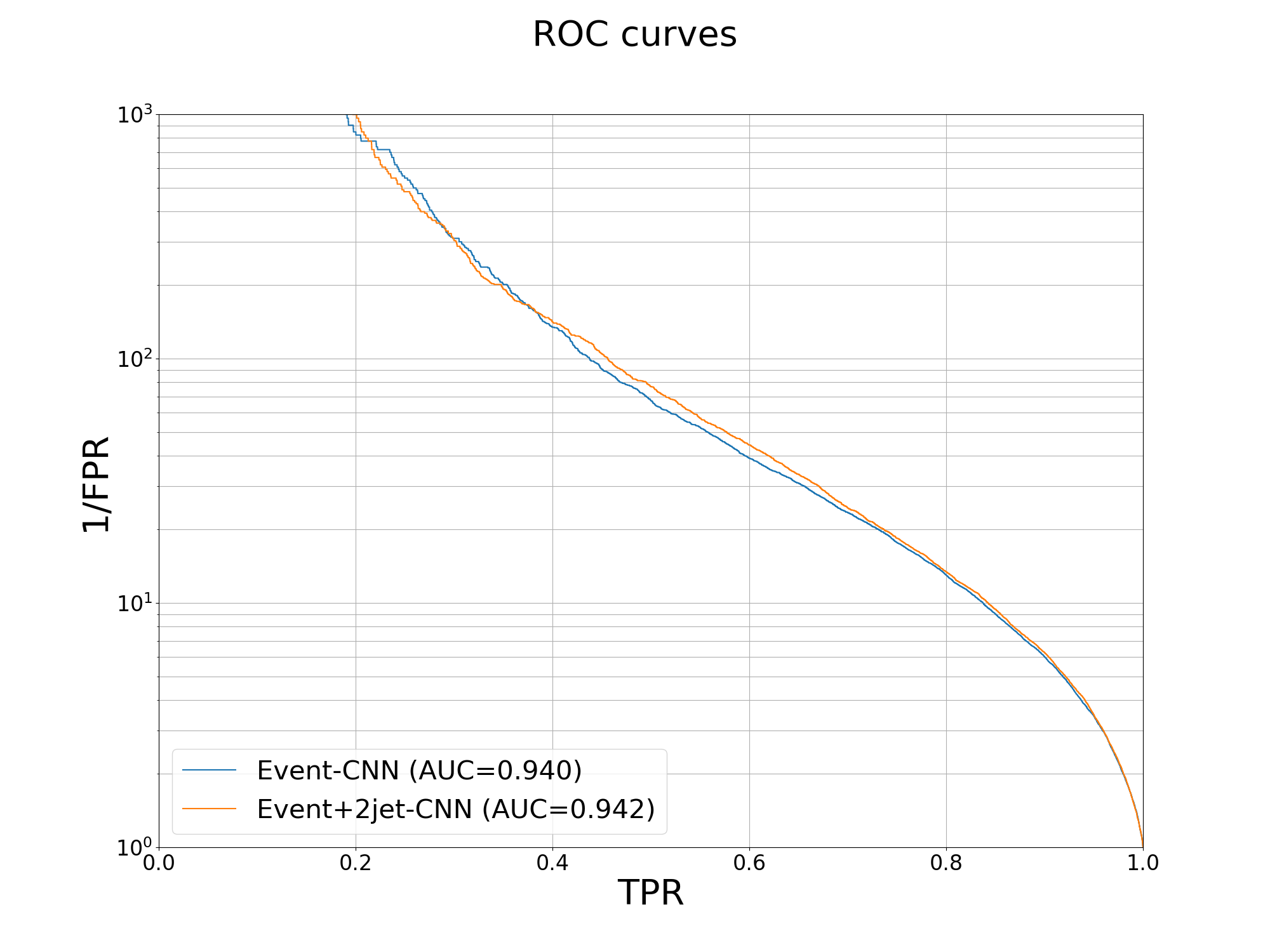}
    \caption{ROC curves of the event-CNN and event+2jet-CNN.}
    \label{fig:evt_ROC3}
\end{figure}

The ROC curves of our original event-CNN and this event + 2jet-CNN are plotted in Fig.~\ref{fig:evt_ROC3}. The curves almost overlap with each other and the AUCs are very similar, indicating that the event-CNN has already captured useful information for classification. The additional high-resolution jet images do not provide significant extra help to the performance. 

\section{Dipole shower in \textsc{Pythia}}
\begin{figure}[htb]
    \centering
    \includegraphics[scale=0.5]{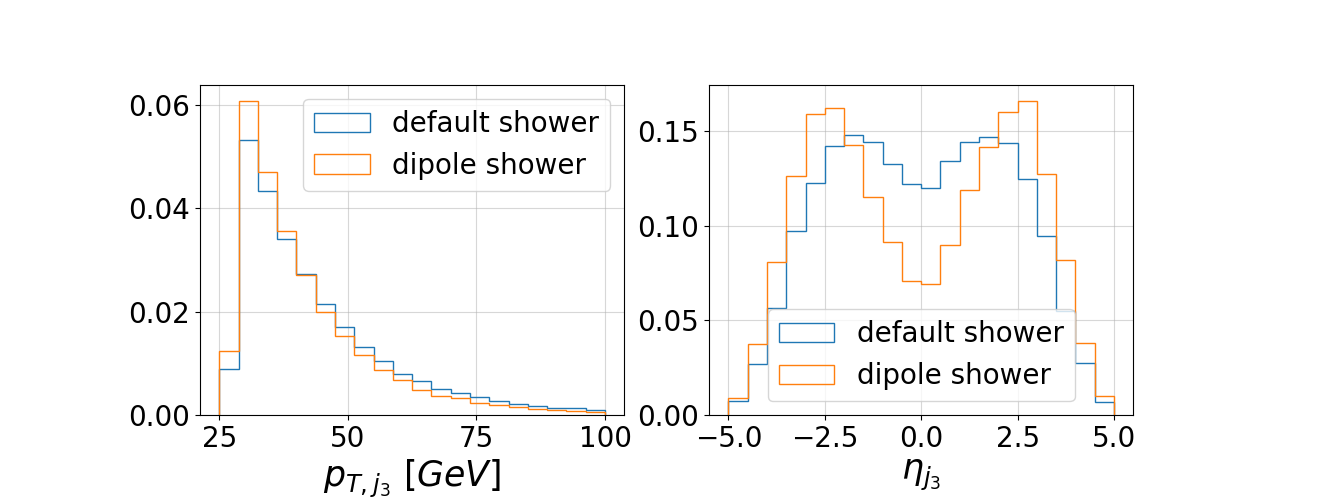}
    \captionsetup{justification=raggedright,singlelinecheck=false}
    \caption{Distributions of the transverse momentum and pesudo-rapidity of the third jet coming from the VBF events passing the VBF pre-selection. All histograms are normalized so that each area under the curve is one.}
    \label{fig:shower_j3var}
\end{figure}

In this section, we present the impact of different \textsc{Pythia} shower schemes applying to the VBF process. In contrast to the local dipole recoil scheme~\cite{Cabouat:2017rzi}, which is the shower scheme we adopt in our study, we also show the kinematic distributions and training results based on samples showered by the default shower scheme in \textsc{Pythia}. As investigated in previous works~\cite{Hoche:2021mkv,Jager:2020hkz,Konar:2022bgc}, the default \textsc{Pythia} shower depicts the emission of additional jets in VBF poorly in the central region. This can be seen in Fig.~\ref{fig:shower_j3var}, which shows an apparent discrepancy on the pseudo-rapidity distribution of the third jet.

\begin{figure}[htb]
    \centering
    \includegraphics[scale=0.25]{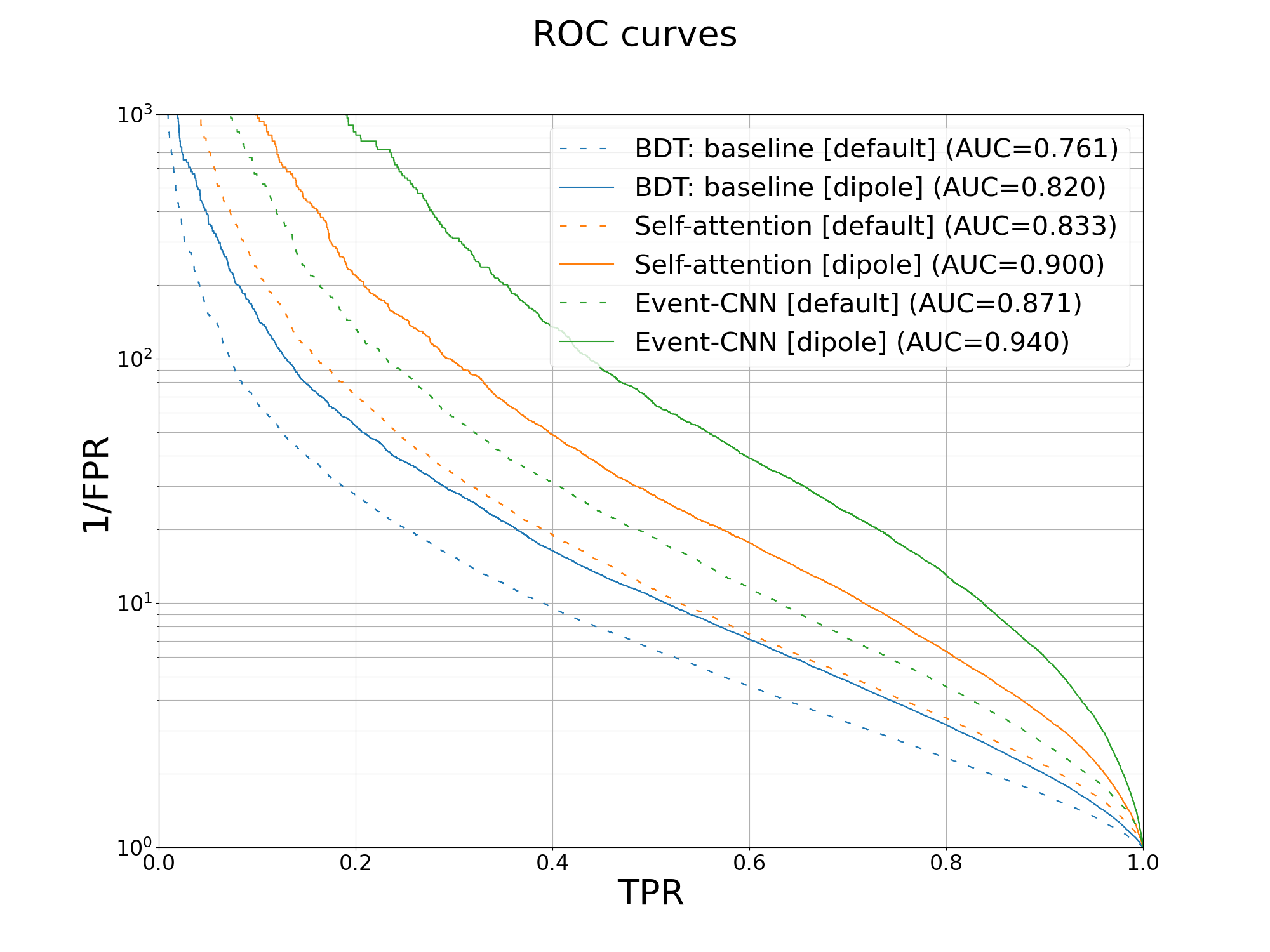}
    \captionsetup{justification=raggedright,singlelinecheck=false}
    \caption{ROC curves of several event-level classifiers in different shower schemes for VBF.}
    \label{fig:evt_ROC6}
\end{figure}

Moreover, the classification performance is severely affected by the choice of the shower scheme. In Fig.~\ref{fig:evt_ROC6}, the performance of each classifier is improved when changing from the default shower to the local dipole shower. The reason of the lower AUC of the default shower is that it produces an incorrect pattern of the emission of the additional jets, which leads to the similarity between VBF and GGF. However, as Fig.~\ref{fig:shower_j3var} shows, the emission issue is solved by the local dipole recoil shower, and hence the performance becomes improved in the local dipole recoil scheme.

\newpage

\bibliographystyle{apsrev4-1}
\bibliography{references}

\end{document}